\documentclass[aps,arxiv,preprint,superscriptaddress,groupedaddress]{revtex4-1}  
\usepackage{graphicx}
\usepackage{dcolumn}
\usepackage{bm}
\usepackage{amsmath}
\usepackage{amssymb}
\usepackage{mathtools}

\usepackage{float}

\usepackage[version=3]{mhchem}

\newlength{\singlecol}
\newlength{\doublecol}
\usepackage{microtype}

\newcommand{\figref}[2][]{\hyperref[#2]{Figure~\ref{#2}#1}}
\newcommand{\tabref}[2][]{\hyperref[#2]{Table~\ref{#2}#1}}

\begin{document}

\title{Atomic-scale factors that control the rate capability of
  nanostructured amorphous Si for high-energy-density batteries}

\author{Nongnuch Artrith}
\email{nartrith@atomistic.net}
\affiliation{%
  Department of Materials Science and Engineering,
  University of California, Berkeley, CA, USA}
\author{Alexander Urban}
\affiliation{%
  Department of Materials Science and Engineering,
  University of California, Berkeley, CA, USA}
\author{Yan Wang}
\affiliation{%
  Department of Materials Science and Engineering,
  Massachusetts Institute of Technology, Cambridge, MA, USA}
\author{Gerbrand Ceder}
\email{gceder@berkeley.edu}
\affiliation{%
  Department of Materials Science and Engineering,
  University of California, Berkeley, CA, USA}
\affiliation{%
  Materials Science Division, Lawrence Berkeley National
  Laboratory, Berkeley, CA, USA}
\date{\today}

\begin{abstract}
  Nanostructured Si is the most promising high-capacity anode material
  to substantially increase the energy density of Li-ion batteries.
  Among the remaining challenges is its low rate capability as compared
  to conventional materials.
  To understand better what controls the diffusion of Li in the
  amorphous Li-Si alloy, we use a novel machine-learning potential
  trained on more than 40,000 \emph{ab-initio} calculations and
  nanosecond-scale molecular dynamics simulations, to visualize for the
  first time the delithiation of entire LiSi nanoparticles.
  Our results show that the Si host is not static but undergoes a
  dynamic rearrangement from isolated atoms, to chains, and clusters,
  with the Li diffusion strongly governed by this Si rearrangement.
  We find that the Li diffusivity is highest when Si segregates into
  clusters, so that Li diffusion proceeds via hopping between the Si
  clusters.
  The average size of Si clusters and the concentration range over which
  Si clustering occurs can thus function as design criteria for the
  development of rate-improved anodes based on modified Si.
\end{abstract}

\maketitle


Lithium-ion batteries have enabled tremendous technological advances in
consumer electronics and electric vehicles during the past
decade~\cite{jacs135-2013-1167}, but to keep up with the growing demand
for portable energy storage improved electrode materials with greater
energy densities will be required.
Over the last years, nanostructured Si has emerged as a promising
high-capacity alternative to conventional graphite-based
anodes~\cite{am25-2013-4966, aem4-2013-1300882, jmca1-2013-9566}.
However, while nanoscaling has allowed to overcome mechanical
limitations such as fracturing~\cite{am27-2015-1526}, the
rate-capability of silicon anodes is 2--4~orders of magnitude
lower~\cite{ssi180-2009-222, mcp120-2010-421, jpcc116-2012-1472} than
graphite's~\cite{jps195-2010-7904} owing to low Li mobility in the
\ce{LiSi} alloy, which considerably limits the power density.
A detailed understanding of the nature of the relevant \ce{Li_xSi}
phases and the structural factors that control Li transport in the LiSi
alloy is required for the rational design of Si-based anodes with
improved Li mobility.

The lithiation of crystalline Si (c-Si) and the rich phase diagram of
the LiSi alloy have been thoroughly investigated both by
experiment~\cite{jssc37-1981-271, am51-2003-1103, esl7-2004-93,
  jes154-2007-156, am25-2013-4966, nl13-2013-758, nl13-2013-709,
  nc5-2014-3217} and simulation~\cite{cjp87-2009-625, jes156-2009-454,
  jes157-2010-392, jpcc115-2011-2514, jacs134-2012-14362,
  nl13-2013-2011, nl14-2014-4065, jpcc119-2015-3447}.
Lithiation of c-Si at room temperature leads to the formation of an
amorphous \ce{Li_xSi} phase (a-\ce{Li_xSi}) that crystallizes below
$\sim$50~mV vs.~\ce{Li+/Li} in the \ce{Li15Si4}
structure~\cite{am51-2003-1103, esl7-2004-93, jes154-2007-156,
  am25-2013-4966}.
The \hbox{a-\ce{Li_xSi}} and the \hbox{c-\ce{Li15Si4}} phases are
metastable, but the crystalline ground states,
c-\ce{Li8Si8}~\cite{jssc173-2003-251},
c-\ce{Li12Si7}~\cite{cb119-1986-3576},
c-\ce{Li7Si3}~\cite{zm71-1980-357}, c-\ce{Li13Si4}~\cite{zn30-2014-10},
and c-\ce{Li21Si5}~\cite{jssc70-1987-48}, are only observed after
annealing at higher temperatures~\cite{jssc37-1981-271}.
After the first lithiation-delithiation cycle, lithium insertion and
extraction only involves a-\ce{Li_xSi} phases with intermediate
compositions.
Computational studies have been important to understand the mechanism of
Li insertion into crystalline Si~\cite{jacs134-2012-14362,
  nl13-2013-2011, jpcc119-2015-3447} and the resulting
amorphization~\cite{cjp87-2009-625, jes156-2009-454, jes157-2010-392},
to describe the topology of a-\ce{Li_xSi} phases~\cite{nl14-2014-4065}
and to relate the formation of \ce{Si-Si} bonds to the mechanical
properties of the material~\cite{nl11-2011-2962}.
However, simulating Li transport through realistic amorphous LiSi
structures requires large structure models limiting first-principles
methods to molten \ce{LiSi} phases at high temperatures and short time
scales below 100~ps~\cite{jpcc115-2011-2514, jps263-2014-252}, as the
extensive sampling required to reach the time scales on which Li
transport occurs at room temperature is not presently feasible.
Additionally, the lack of long-range order prevents the use of
experimental diffraction techniques for structure characterization, so
that only indirect experimental information about the amorphous LiSi
structures is available.

To shed light on the Li diffusion and extraction mechanism in
nanostructured amorphous LiSi anodes, we employed advanced computational
techniques to simulate the delithiation of entire \ce{LiSi}
nanoparticles on the atomic scale.
Our simulations are based on a combination of first-principles
calculations and long-time-scale molecular dynamics (MD) simulations
using highly accurate machine-learning potentials that have been
carefully validated.
The results elucidate how the clustering of Si atoms affects the Li
diffusivity and identify structural motifs that are beneficial for Li
transport, thereby providing a guideline for the design of Si-based
anodes with improved rate capability.


\begin{figure*}[tbp]
  \centering
  \includegraphics[width=\textwidth]{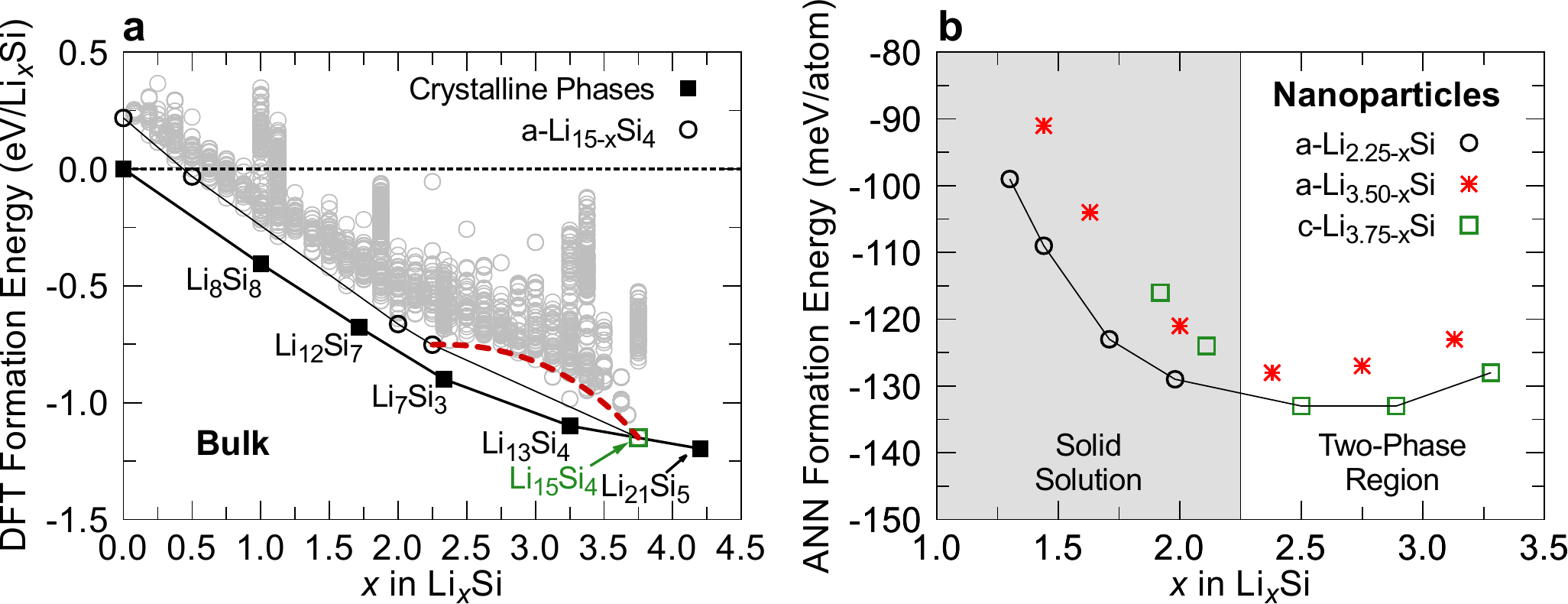}
  \vspace{-0.5\baselineskip}
  \caption{\label{fig:formation-energies}%
    \textbf{Crystalline \ce{Li15Si4} segregates from amorphous
      \ce{Li_xSi} for $\bm{x>2.25}$.} \textbf{a},
    Density-functional theory formation energies of the crystalline bulk
    \ce{Li_xSi_y} structures (filled squares) and the amorphous bulk
    \ce{Li_{15-x}Si_{4}} structures samples for the present work (empty
    circles).  The amorphous phase exhibits a two-phase region for
    $x>2.25$ (red dashed line) in which phase separation into
    crystalline \ce{Li15Si4} (empty green square) and a-\ce{Li_{2.25}Si}
    is thermodynamically favored. \textbf{b}, The ANN potential
    formation energies of three nanoparticle models at different stages
    of delithiation after 4~ns equilibration at $T$~=~500\,K.
    Consistent with the bulk phase diagram in \textbf{a}, the most
    stable nanoparticle for compositions \ce{Li_xSi} with $x>2.25$ are
    derived from crystalline \ce{Li15Si4} (empty squares) whereas for
    $x<2.25$ a solid solution behavior is observed (empty circles).}
\end{figure*}

\section{Accurate nanoscale simulations}
To reach the length and time scales required for the simulation of Li
transport through LiSi nanoparticles without sacrificing predictive
accuracy, we developed a state-of-the-art artificial neural network
(ANN) machine-learning potential~\cite{prl98-2007-146401,
  cms114-2016-135, ArtrithUrbanCeder2016} that was trained to reproduce
density-functional theory (DFT) energies based on a reference set of
40,653~DFT bulk, cluster, and surface structures with different
compositions.
Similar approaches have recently been successfully applied to the
modeling of amorphous Si and LiSi alloy phases~\cite{Artrith2018,
  prb97-2018-094106, jpcl9-2018-2879, Bernstein2018-arxiv}.
Our ANN potential achieves a root mean squared error of 7.7~meV/atom and
a mean absolute error of 5.9~meV/atom relative to the DFT reference
energies as determined based on an independent validation set of
4,516~randomly selected structures that were not used for training
(\textbf{Fig.~\ref{SI-fig:ANN-error-distribution}}).
Supplementary \textbf{Fig.~\ref{SI-fig:hull-all-structures}} and
\textbf{Fig.~\ref{SI-fig:voltage-profile}} show that the formation
energies of \ce{Li_xSi} structures and the corresponding voltages
predicted by the ANN potential are in excellent agreement with their DFT
references.
All DFT ground states are correctly reproduced by the ANN potential.
Additionally, the ANN potential predicts Li diffusion barriers in
excellent quantitative agreement with DFT, as we confirmed with
transition path calculations using the \emph{nudged elastic band}
method~\cite{Jonsson1998, jcp113-2000-9901}, and thus enables reliable
molecular dynamics simulations (\textbf{Fig.~\ref{SI-fig:NEB}} to
\textbf{Fig.~\ref{SI-fig:MD-ANN-vs-DFT}}).
Further information about the parameters of the DFT calculations and the
ANN potential can be found in the methods section.


\section{Amorphous $\text{Li}_x\text{Si}$ bulk phase diagram}
For the generation of amorphous \ce{Li_xSi} bulk phases, a supercell of
the c-\ce{Li15Si4} structure with composition \ce{Li480Si128} was
computationally delithiated using a combination of the ANN potential and
a genetic algorithm~\cite{Artrith2018, Lacivita2018}.
The protocol used for the generation of bulk a-\ce{Li_xSi} structures is
detailed in the methods section, and a comparison with measured pair
distribution functions from the literature confirms that the structures
agree well with experiment (Supplementary
\textbf{Fig.~\ref{SI-fig:pdf-Li480Si128}} and related discussion in the
Supplementary Information).
The formation energies of the resulting metastable amorphous structures,
recomputed with DFT, are shown in
\textbf{Fig.~\ref{fig:formation-energies}a} together with the
ground-state hull defined by the crystalline LiSi phases.
As seen in the figure, the amorphous phase diagram exhibits a two-phase
region between \ce{Li_{2.25}Si} and \ce{Li_{3.75}Si} (=~\ce{Li15Si4}) in
which phase separation into the end-members is favored.
For compositions \ce{Li_xSi} with $x<2.25$ a solid-solution behavior is
observed, i.e., each composition exhibits a formation energy on or close
to the amorphous convex hull.
A similar phase diagram has been previously reported based on
simulations using a smaller cell size~\cite{jpcc117-2013-18796}.
Bulk structure models are not appropriate to gain insight into the
delithiation mechanism in actual LiSi nanostructures, but the bulk phase
diagram provides a basis for the construction of reasonable nanoparticle
models for subsequent investigation.


\begin{figure*}[tbp]
  \centering
  \includegraphics[width=\textwidth]{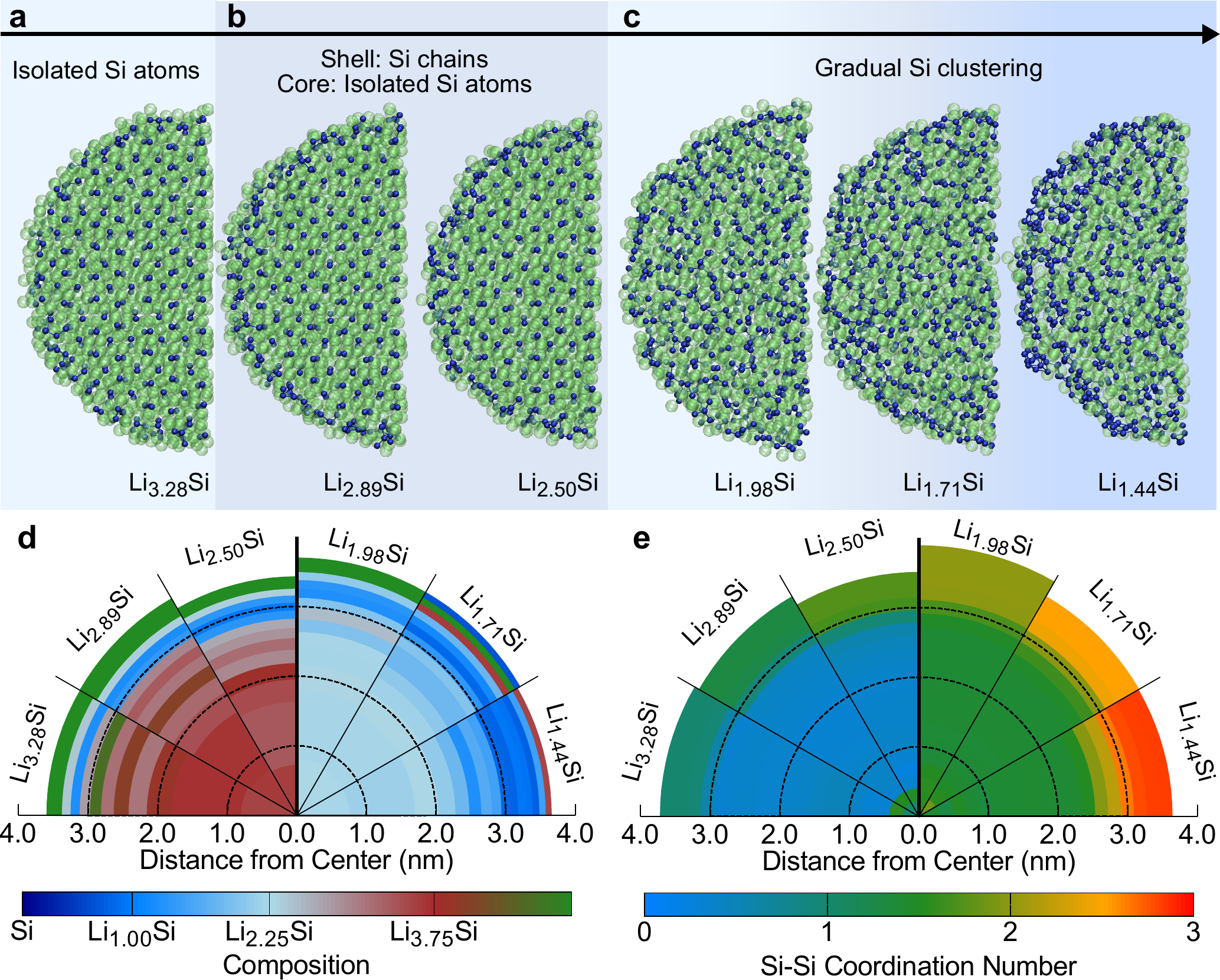}
  \caption{\label{fig:NP12000}%
    \textbf{Si atoms segregate into clusters
      and chains during the delithiation of LiSi nanoparticles with
      12,000 atoms.}  \textbf{a--c}, Cross sections of nanoparticle
    structures corresponding to six stable phases of
    Fig.~\ref{fig:formation-energies}b with decreasing lithium content.
    The nanoparticles shown in \textbf{a} and \textbf{b} were obtained
    by delithiation of a c-\ce{Li_{3.75}Si} (\ce{Li_{9429-x}Si2571})
    nanoparticle extracted from the \ce{Li15Si4} crystal structure.  Li
    was sequentially removed from the particle surface followed by an
    equilibration with molecular dynamics simulations over 4~ns at 500~K
    at each composition.  The structures in panel \textbf{c} were
    equivalently derived from a nanoparticle with original composition
    \ce{Li_{2.25}Si} (\ce{Li_{8310-x}Si3690}).  Li and Si atoms are
    colored green and blue, respectively.  \textbf{d},~Internal
    compositions of the nanoparticles as function of the distance from
    the particle center.  The fully lithiated composition
    (\ce{Li_{3.75}Si}) is colored red-brown, green regions are Li rich,
    and blue regions indicate relative delithiation.  \textbf{e},~Si-Si
    coordination number as function of the distance from the particle
    center for the same nanoparticles.  The data shown in \textbf{d} and
    \textbf{e} was obtained by averaging over the final 50~ps of 4~ns
    molecular dynamics trajectories at 500~K.}
\end{figure*}
\section{Computational delithiation of $\text{LiSi}$ nanoparticles}
To ensure the convergence of our simulations with particle size, we
considered particles with different diameters between 2~and~8~nm and
find that the structural motifs are converged for particle diameters
between 6~to~8~nm (6,000--12,000~atoms).
The following discussion is therefore based on the simulations of
particles with diameters of around 8~nm corresponding to
$\sim$12,000~atoms in their fully lithiated composition.
\emph{In~situ} NMR measurements have previously revealed that the
delithiation mechanism depends on the initial degree of
lithiation~\cite{nc5-2014-3217}.
This observation also agrees with the phase diagram in
\textbf{Fig.~\ref{fig:formation-energies}a}.
Therefore, we constructed nanoparticle structures by extraction from
\ce{Li_{15-x}Si4} bulk structure models with three different
compositions, \hbox{c-\ce{Li_{3.75}Si}} (=~\ce{Li15Si4}),
\hbox{a-\ce{Li_{3.50}Si}}, i.e., within the bulk two-phase region, and
\hbox{a-\ce{Li_{2.25}Si}}, i.e., the fully lithiated end-member of the
solid-solution phase.

At each delithiation step Li atoms were extracted from the particle
surface, which was followed by a thermal equilibration over 4~ns at
$T$~=~500~K.
To parallelize some of the computational effort, nanoparticle models of
the next lower Li content were constructed based on the structure after
1~ns equilibration, as we generally found that sufficient Li atoms had
migrated to the particle surface to allow for further
delithiation.
The potential energy during the MD simulations
(\textbf{Fig.~\ref{SI-fig:MD-equi}}) is generally converged after 2~ns
showing that the structure of the nanoparticles has reached thermal
equilibrium.

The formation energies of the various nanoparticle structures relative
to elemental Li and Si are shown in
\textbf{Fig.~\ref{fig:formation-energies}b}.
For compositions \ce{Li_xSi} with $x<2.25$ the nanoparticle model based
on the bulk a-\ce{Li_{2.25}Si} structure is most stable, for higher Li
contents the delithiated crystalline \ce{Li_{3.75-x}} particles are
lowest in energy.
The nanoparticle structures based on the a-\ce{Li_{3.50}Si} is never
stable, in accordance with the bulk phase diagram.

\textbf{Figure~\ref{fig:NP12000}a--c} shows cross sections of six
nanoparticle structures at different stages of delithiation that are on
the ground-state hull in \textbf{Fig.~\ref{fig:formation-energies}b}.
As seen in \textbf{Figure~\ref{fig:NP12000}a--b}, Si chains and clusters
first form in the outer regions of the particle upon delithiation, while
the Si atoms in the core of the particle initially remain isolated.
Continued Li extraction leads to the growth of the clustered phase from
the particle surface towards the center.
Once the overall composition \ce{Li_{2.25}Si} has been reached, the
delithiation proceeds via a single-phase mechanism
(\textbf{Figure~\ref{fig:NP12000}c}), so that the length of the Si
chains within the entire cluster increases upon Li extraction.
Eventually, larger Si clusters form and as the overall composition
reaches $\sim$\ce{Li_{1.3}Si} the structure is dominated by Si clusters
and chains.

Note that the predicted delithiation mechanism is fully consistent with
previously reported experimental evidence~\cite{nl13-2013-758,
  nl13-2013-709, nc5-2014-3217}.
However, our simulations allow for the first time to unambiguously
characterize the atomic structures of all phases occurring during the
delithiation.
For this purpose, the internal composition and the nature of the Si
clustering as function of the distance from the particle center at
different stages of delithiation is shown in
\textbf{Figure~\ref{fig:NP12000}d--e}.
This data was determined by computing histograms for the final 50~ps of
the corresponding 4~ns MD trajectories using spherical bins with equal
volume.

As seen in \textbf{Figure~\ref{fig:NP12000}d}, during the initial stages
of delithiation the core of the particle remains close to the original
\ce{Li_{3.75}Si} composition.
Further, the Li concentration does not decrease monotonously from the
particle center towards the surface, but instead reaches a local minimum
at a distance of around three fourth of the particle radius.
This is the phase that grows during delithiation in the two-phase
regime.

Once the Li content drops below 2.25, a more homogeneous delithiation is
energetically favored, so that the Li content in nanoparticles with
compositions \ce{Li_{1.98}Si}, \ce{Li_{1.71}Si}, and \ce{Li_{1.44}Si}
decreases continuously from the particle center to the surface.

We stress that until an overall composition of \ce{Li_{1.71}Si} has been
reached, Li is the predominant species at the particle surface after
equilibration, indicating that the simulated delithiation rate is not
too fast.
For the particle with lower Li content (\ce{Li_{1.44}Si}) the surface
contains some Si, evidencing that the overall composition has not
equilibrated on the time scale of our simulation.
%
%
The Li content can therefore be expected to homogenize further on
experimental time scales in the \ce{Li_{1.71}Si} and \ce{Li_{1.44}Si}
compositions.

Note that the driving force for the Li atoms to migrate to the particle
surface in our simulations is the lower Li surface energy as compared to
Si.
In an actual electrochemical cell the lower free energy of oxidation of
Li can be expected to have a similar effect.

The Si clustering behavior can be further quantified by considering the
\ce{Si-Si} coordination number, shown for a bond-length cutoff of
2.7~\AA{} in \textbf{Figure~\ref{fig:NP12000}e}.
As seen in the figure, Si atoms in the center of the particle remain
mostly isolated as long as the composition in the particle core remains
close to \ce{Li_{3.75}Si}.
The Si clustering in \ce{Li_xSi} with $2.25<x<3.75$ is more pronounced
near the surface as evidenced by the higher coordination numbers.
Based on the coordination number, the Si clustering is similar to the
one in the \ce{Li_{2.25}Si} phase with an average coordination 1.0--1.5,
corresponding to Si dimers and trimer chains.
Further delithiation to \ce{Li_{2.25-x}Si} compositions results in a
more homogeneous clustering behavior with continuously increasing
coordination number from the particle center to the surface.

\section{$\text{Li}$ diffusivity in the relevant $\text{LiSi}$ phases}
The nanoparticle simulations resolve the structure of the LiSi alloy at
different states of charge and identify the particular importance of the
\ce{Li_{3.75}Si} (\ce{Li15Si4}) and \ce{Li_{2.25}Si} phases that exhibit
fundamentally different Si clustering behavior.
To understand how the Si clustering affects the Li diffusion mechanism
and the Li diffusivity, we carried out 5~ns long MD simulations of bulk
structures with the initial nanoparticle compositions (\ce{Li_{3.75}Si},
\ce{Li_{3.50}Si}, and \ce{Li_{2.25}Si}) and with the composition of a
further delithiated phase with increased Si clustering
(\ce{Li_{1.00}Si}).
In light of the low measured Li diffusivities on the order of
10$^{-10}$--10$^{-14}$~cm$^2$s$^{-1}$~\cite{ssi180-2009-222,
  jpcc113-2009-11390, mcp120-2010-421, jpcc116-2012-1472,
  nl13-2013-1237, aem5-2014-1401627} (see also
Table~\ref{SI-tab:diffusivity}), elevated temperatures (400~K, 500~K,
600~K, 700~K, 800~K, 900~K, 1000~K, 1100~K, and 1200~K) were chosen, so
that sufficient Li hopping events occur on the timescale of the
simulation.
Room temperature diffusivities were obtained by Arrhenius extrapolation,
which has proven reliable for similar
applications~\cite{nc7-2016-11009}.
Since the lowest melting temperature of any \ce{Li_xSi} composition is
865\,K for \ce{Li_{0.75}Si}~\cite{am51-2003-1103}, i.e., below the
maximal temperatures in our simulations, we confirmed that for
simulations above 800\,K the time scale was too short for melting to
occur, allowing us to capture diffusion in the solid state.

\begin{table}[tb]
  \caption{\label{tab:diffusivity}%
    \textbf{Li diffusivity determined by molecular dynamics
      simulations.}  Calculated activation energies $E_{\textup{a}}$ for
    Li diffusion in amorphous \ce{Li_xSi} and Arrhenius-extrapolated
    diffusivity $D$ at room temperature.}
\renewcommand{\arraystretch}{1.0}
\begin{tabular}{l@{\quad}l@{\quad}l}
\hline\hline
  \multicolumn{1}{c}{$x_{\textup{Li}}$} &
  \multicolumn{1}{c}{$E_{\textup{a}}$ (eV)}  &
  \multicolumn{1}{c}{$D$ (cm$^2$s$^{-1}$)} \\
\hline
 0.75 & 0.789 $\pm$ 0.022 & (1.154 $\pm$ 1.084)$\times{}$10$^{-14}$ \\
 1.00 & 0.500 $\pm$ 0.015 & (5.986 $\pm$ 3.722)$\times{}$10$^{-11}$ \\
 2.25 & 0.483 $\pm$ 0.010 & (9.607 $\pm$ 3.805)$\times{}$10$^{-11}$ \\
 3.50 & 0.682 $\pm$ 0.027 & (3.820 $\pm$ 4.691)$\times{}$10$^{-13}$ \\
\hline\hline
\end{tabular}
\end{table}
The computed room-temperature Li diffusivities and the corresponding
activation energies shown in \textbf{Table~\ref{tab:diffusivity}} vary
strongly with the composition and span a large range.
In the fully lithiated \ce{Li_{3.75}Si} structure, the diffusivity is
lowest and is indistinguishable from zero in our MD simulations.
%
Delithiation to \ce{Li_{3.50}Si} results in an improved but still low Li
diffusivity of $\sim$10$^{-13}$\,cm$^{2}$s$^{-1}$.
A significantly higher diffusivity of $>$10$^{-11}$\,cm$^{2}$s$^{-1}$ is
found for the clustered amorphous \ce{Li_{2.25}Si} and \ce{Li_{1.00}Si}
compositions.
At lower Li contents Si forms extended networks throughout the structure
impeding Li diffusion, and the Li diffusivity in the \ce{Li_{0.75}Si}
composition is only $\sim$10$^{-14}$\,cm$^{2}$s$^{-1}$.
Note that the predicted diffusivities for all delithiated compositions
fall into the experimentally measured range, though the variation of the
diffusivity with the state of charge depends on the morphology and size
of the electrode particles and the prevailing delithiation mechanism
(single-phase or two-phase).

\begin{figure*}
  \centering
  \vspace*{-\baselineskip}%
  \includegraphics[width=\linewidth]{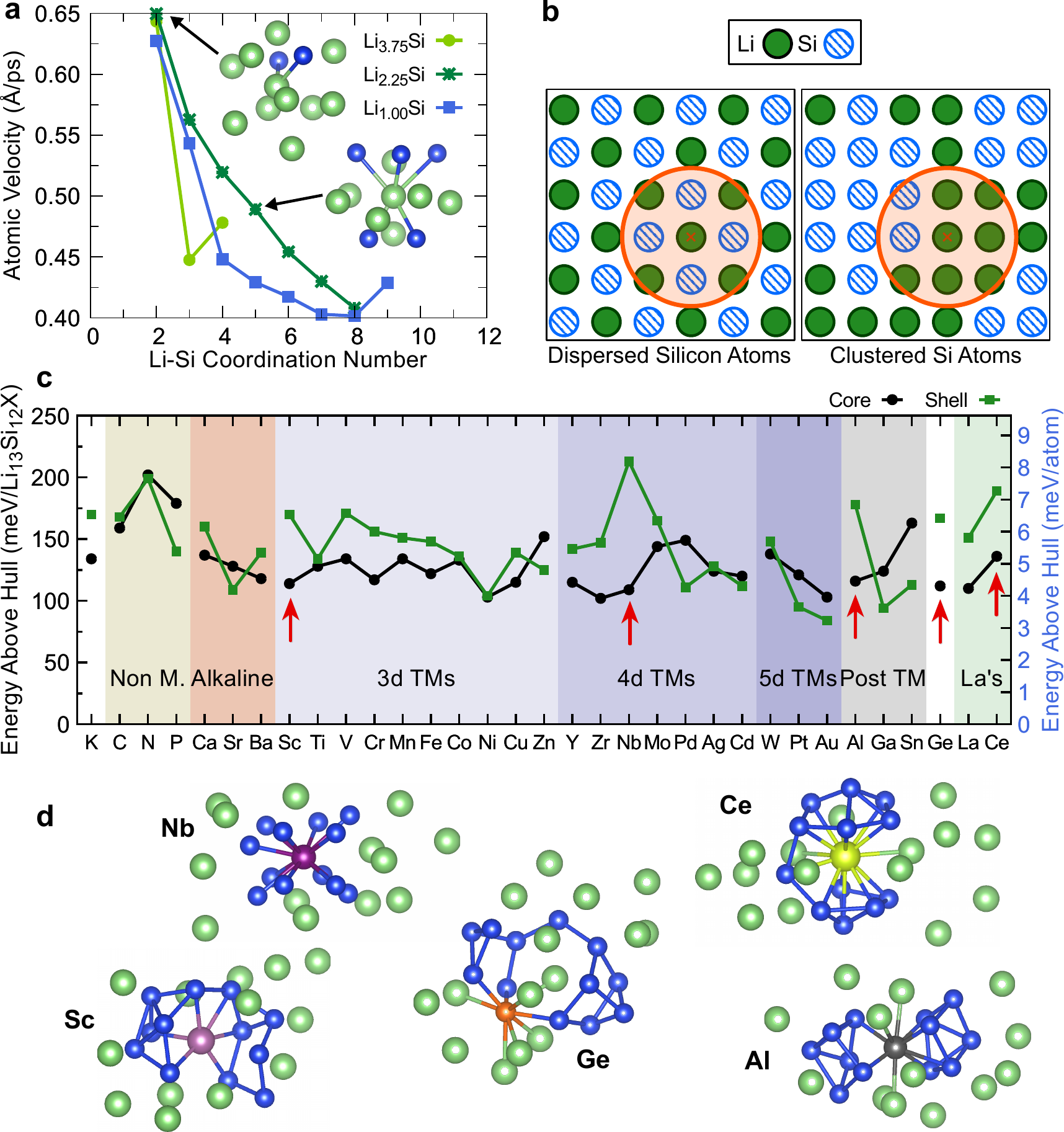}
  \caption{\label{fig:diffusion-mechanism}%
    \textbf{Si segregation is beneficial for Li diffusion.} \textbf{a},
    The average velocity of Li atoms as function of the number of Si
    atoms within 3.5~\AA{} around them.  The velocity decreases with
    increasing number of Si atoms. Two representative Li environments
    with two and five Si atoms are shown in the insets. \textbf{b},
    Schematic showing that the average Li-Si coordination number
    decreases when Li and Si segregate into separate
    domains. \textbf{c}, Decomposition energies of Si clusters embedded
    in Li doped with a third species X either on the surface (shell;
    green squares) or inside (core; black circles) of the Si cluster
    with composition \ce{Li13Si12X}.  The five systems for which the
    core structure is most stabilized are indicated with red arrows.
    \textbf{d}, Optimized geometries of the five core-doped
    \ce{Li13Si12X} compositions highlighted in panel \textbf{c}.}
\end{figure*}
\section{Understanding $\text{Li}$ diffusion in the presence of $\text{Si}$ clusters}
The drastic change of the Li diffusivity with the composition indicates
a strong dependence of the Li transport on the distribution of Si atoms
throughout the material.
By visualizing the Li trajectories we observed that Li atoms spend much
time in the vicinity of Si clusters and only occasionally hop between
different clusters.

To better understand the extent to which the Li mobility in the
different \ce{Li_xSi} phases is determined by the presence of Si
clusters, we analyzed 5~ns long trajectories with respect to the effect
of coordinating Si atoms on the diffusion velocity of all Li atoms.
For this analysis, we assigned to each Li atom all Si atoms within a
cutoff range of 3.5~\AA{} over the course of the MD simulation, which
allows us to determine the average Li diffusion velocity as a function
of the Si coordination number.
The results for those coordination numbers that occur with a frequency
of at least 1\% are shown in
\textbf{Fig.~\ref{fig:diffusion-mechanism}a} for the three compositions
\ce{Li_{3.75}Si}, \ce{Li_{2.25}Si}, and \ce{Li_{1.00}Si}.

The insights from the coordination analysis are \textbf{(i)}~that the Li
diffusion velocity decreases steeply with increasing Si coordination
number, indicating a strong Li-Si interaction.
\textbf{(ii)}~The highest atomic velocity for Li atoms coordinated by
2~Si atoms is nearly identical for the three considered compositions.
\textbf{(iii)}~The differences in overall Li diffusivity instead arise
from the different distribution of coordination numbers in the three
compositions (\textbf{Fig.~\ref{SI-fig:cn-vs-composition}}):
During the simulation of the crystalline \ce{Li_{3.75}Si} structure Li
is symmetrically coordinated by exactly 3~Si atoms more than 80\% of the
time, locking the Li atom in place.  Lower coordinations occur with only
close to 1\% probability.
However, in the amorphous \ce{Li_{2.25}Si} and \ce{Li_{1.00}Si}
compositions the probability of a coordination number of 2~is increased
relative to the \ce{Li_{3.75}Si} phase even though the average Si
coordination number is greater because of the reduced Li content.

The \emph{increased} amount of Li atoms with low Si coordination number
at \emph{decreased} Li content can be understood in the context of Si
clustering.
As schematically shown in \textbf{Fig.~\ref{fig:diffusion-mechanism}b},
Si segregation in form of clustering reduces the average number of Si
atoms around Li.
Since Li diffuses fast only when it detaches from a Si cluster or chain,
the Li diffusion mechanism is akin to Li hopping between Si clusters.
A maximal overall Li diffusivity would therefore be expected for a
structure with perfect Si segregation into large clusters.

The diffusivity $D$ in such a \emph{discrete} lattice site hopping model
can be expressed as~\cite{pms55-2010-61, acr46-2013-1216}
\begin{align*}
  D = \rho\lambda^{2}\Gamma
\end{align*}
where $\Gamma$ is the hopping rate between Si clusters, and $\lambda$ is
the mean distance between the Si clusters, and the geometry factor
$\rho$ depends on the spatial arrangement of the clusters.
Hence, the diffusivity increases approximately quadratically with the
mean cluster distance $\lambda$.
The hopping rate $\Gamma$ can be taken to be constant as it is
determined by the probability of Li atoms to be trapped near a Si
cluster and thus mainly depends on the Li-Si interaction strength.
Assuming that the density of the LiSi alloy is independent of the
arrangement of the Si atoms, the mean free distance between Si clusters
in a close-packed structure is proportional to the average cluster
diameter.
This important finding explains why Si clustering is beneficial for Li
diffusion, as it increases the mean free distance between Si atoms
compared to homogeneously dispersed isolated Si atoms.
Si clustering can thus function as a design criterion for the
engineering of improved Si-based anodes, for example, as composite
materials~\cite{jps246-2014-167, jmca2-2014-1257} or by co-alloying or
doping of other chemical species.
It is known from other alloys that even small dopant concentrations can
promote the segregation of some of the
constituents~\cite{jms38-2003-1203}.

To demonstrate that doping can in fact improve the Si clustering in the
LiSi alloy, we evaluated the DFT energy of doped model Si clusters.
Using the Materials Project open database~\cite{aplm1-2013-011002}, we
identified potential dopant species that are known to form compounds
with Si but not with Li.
Other common species in Si compounds such as C were also considered,
giving a total of 33~dopant species.

To determine in a first approximation whether the dopants would favor Si
clustering, an initial structure of icosahedral 13-atom Si clusters
surrounded by 13~Li atoms in a periodic close-packed arrangement was
constructed.
From this ideal cluster model, two structures were derived for each
dopant species (\textbf{Fig.~\ref{SI-fig:ideal-doped-clusters}}): One
structure in which the central Si atom of the \ce{Si13} cluster is
replaced by the dopant atom (\emph{core} model) and one structure in
which one of the other 12~equivalent Si atoms is replaced (\emph{shell}
model).
Both core and shell structures and the lattice parameters were optimized
with DFT calculations, and the resulting formation energies are shown in
\textbf{Fig.~\ref{fig:diffusion-mechanism}c}.

As seen in the figure, the core doping is favored by 21~out of
33~dopants, and the energy above the hull is generally below
10~meV/atom.
The largest stabilization of the core-doped structure is observed for
doping with Sc, Nb, Al, Ge, and Ce (red arrows in
\textbf{Fig.~\ref{fig:diffusion-mechanism}c}).
The optimized core-doped structures for those dopant species are shown
in \textbf{Fig.~\ref{fig:diffusion-mechanism}d}, and the most
well-defined clustering occurs for Nb and Ce.
Note that the Materials Project database contains Si compounds with both
Nb and Ce but does not contain any stable Li-Nb or Li-Ce compounds
(\textbf{Fig.~\ref{SI-fig:PD-Nb-Si-Ce-Si}}).
Doping with the 4d transition metal Mo~\cite{ami8-2016-16862} and with
the Lanthanides La~\cite{mse1-2009-12030, jps235-2013-29} and
Gd~\cite{ml130-2014-61} has previously been reported to enhance the
electrochemical properties of the LiSi alloy, further showing that an
improvement by Nb and Ce doping is plausible.

While this simple computational screening should just be considered a
proof of concept, it demonstrates that doping will indeed affect the Si
clustering behavior.
Further investigation will be required to determine whether the doped
phases are stable at operation conditions and to understand the effect
of doping on the Li-Si interaction strength.

Our results give important insight into the unusual delithiation and
diffusion mechanism of Li in Si.
The findings also point out several avenues towards improved Li
transport in Si-based anodes.
The Li diffusivity can be enhanced by favoring the formation of larger
Si clusters over a wider range of \ce{Li_xSi} compositions.
Compositional additives that improve the Si mobility are more likely to
lead to such larger clusters.
Similarly, the Li diffusivity could be improved by increasing the
effective rate by which Li hops between the Si clusters by co-alloying
with other species which preferentially bind to the surface of the Si
cluster and reduce the attraction between Li and these surfaces.
Ideally, the Li-Si surface interaction would be equal to the Li-Li
interaction causing Li to freely hop from cluster to cluster.

\section{Conclusions}
\vspace*{-0.5\baselineskip}
Using newly developed neural network potentials we were able to simulate
the complex diffusion of lithium in crystalline and amorphous silicon
nanoparticles.
We found as Li is extracted from a highly lithiated nanoparticle, Si
segregates into clusters and chains affecting the Li transport.
Li-rich domains lead to high Li mobility whereas an increasing
coordination by Si slows down diffusion as Li mostly diffuses around Si
clusters.
Owing to this mechanism, the Li diffusivity in the clustered phase is
with around~10$^{-11}$\,cm$^{2}$s$^{-1}$ several orders of magnitude
higher than the diffusivity in phases with isolated Si atoms.
Increasing the cluster size or reducing the Li-Si surface interaction
would all help in increasing the Li transport through silicon.
This finding indicates that the low rate-capability of anodes based on
amorphous Si is not inevitable but could in fact be avoided if the
formation of Si clusters at higher lithium concentrations could be
promoted.


\section{Acknowledgements}
\vspace*{-0.5\baselineskip}
The authors thank China Automotive Battery Research Institute, Co., Ltd.
and General Research Institute for NonFerrous Metals (GRINM) for
financial support.
This work used the computational facilities of the Extreme Science and
Engineering Discovery Environment (XSEDE), which is supported by
National Science Foundation Grant No. ACI-1053575.
Additional computational resources from the University of California
Berkeley, HPC Cluster (SAVIO) and from the Lawrencium computational
cluster resource provided by the IT Division at the Lawrence Berkeley
National Laboratory (Supported by the Director, Office of Science,
Office of Basic Energy Sciences, of the U.S. Department of Energy under
Contract No. DE-AC02-05CH11231) are also gratefully acknowledged.


\section{Author contributions}
\vspace*{-0.5\baselineskip}
G.C.\ and N.A.\ conceived the project.
G.C.\ supervised all aspects of the research, contributed to the data
analysis and to writing the manuscript.
N.A.\ developed the ANN potential, carried out DFT calculations and MD
simulations, analyzed the data, and authored the manuscript.
A.U.\ provided the genetic algorithm and contributed to the data
analysis and to writing the manuscript.
Y.W.\ performed preliminary DFT calculations and contributed to the
discussion.


\newpage
\bibliographystyle{aipnum4-1}
\bibliography{./Artrith-References.bib}


\appendix
\renewcommand{\thefigure}{S\arabic{figure}}
\renewcommand{\thetable}{S\arabic{table}}
\setcounter{figure}{0}
\setcounter{table}{0}

\section{Methods}

\subsection{DFT calculations}

The baseline method for our simulations is accurate electronic DFT using
the Perdew-Burke-Ernzerhof~\cite{prl77-96-3865, prl78-97-1396}
exchange-correlation functional and projector-augmented wave
pseudopotentials~\cite{prB50-1994-17953}, as implemented in the Vienna
Ab-Initio Simulation Package~\cite{cms6-1996-15, prB54-1996-11169}.
We employed a plane-wave basis set with an energy cutoff of 520 eV for
the representation of the wavefunctions and a gamma-centered k-point
grid with a density of 1000 divided by the number of atoms for the
Brillouin zone integration.
The atomic positions and lattice parameters of all structures were
optimized until residual forces were below 20 meV/\AA{}.

\subsection{Artificial neural network potential}

The atomic energy network package (\ae{}net)~\cite{cms114-2016-135} was
used for the construction and application of the artificial neural
network (ANN) machine-learning potential.
ANN potentials are trained to predict the atomic energy based on the
local atomic environment~\cite{prl98-2007-146401} from which the atomic
forces can be evaluated by analytical differentiation.  In this work,
the local atomic environment was described by an orthogonal Chebyshev
basis~\cite{ArtrithUrbanCeder2016}.
The potential is based on ANNs with two hidden layers and each 15~nodes,
using hyperbolic tangent activation functions.
For the ANN training, we employed the limited-memory
Broyden-Fletcher-Goldfarb-Shanno method~\cite{siam16-1995-1190,
  toms23-1997-550}.

\subsection{Molecular dynamics simulations}

All MD simulations were carried out using the Tinker
software~\cite{jocc8-1987-1016} and the ANN potential via an interface
with the \ae{}net package~\cite{cms114-2016-135}.
A time-step of 2~fs was used for the integration of the equation of
motion with the Verlet algorithm, and the canonical ($NVT$) statistical
ensemble was imposed with a Bussi-Parrinello
thermostat~\cite{jcp126-2007-14101}.

\subsection{Delithiation protocol}

Amorphous \emph{bulk} \ce{Li_xSi} structures were generated by
delithiation of a supercell of the \ce{Li15Si4} structure with
composition \ce{Li480Si128}.
A genetic algorithm coupled with the ANN potential was employed to
determine near-ground-state Li-vacancy orderings at each intermediate
\ce{Li_xSi} composition.
A schematic of the procedure is shown in Supplementary
\textbf{Fig.~\ref{SI-fig:GA-schematic}}.
For the \emph{nanoparticle} delithiation, the initial particle was
sequentially delithiated by removing at each step those $N$ Li atoms
furthest away from the particle center followed by a thermal
equilibration by MD simulation over 1~ns at 500~K, where $N$ was chosen
depending on the Li concentration at the particle surface.
The Li removal was followed by another MD equilibration over 3~ns at
500~K (i.e., reaching a total of 4~ns).
In general, $N=1000$ Li atoms were removed during each delithiation
step.
A schematic of the delithiation protocol is shown in Supplementary
\textbf{Fig.~\ref{SI-fig:NP-Delith}}.
A slightly elevated temperature was chosen to allow for sufficient
redistribution of the Li atoms during the equilibration period, as the
simulated delithiation rate is high compared to experimentally
realizable currents.
As seen in Supplementary \textbf{Fig.~\ref{SI-fig:MD-equi}}, the
potential energy during the MD simulations converges within 2~ns, so
that a total equilibration period of 4~ns is generous.

\subsection{Lithium diffusivity in LiSi bulk structures from MD simulations}

\begin{figure}[tbp]
  \hspace*{-8mm}%
  \includegraphics[width=1.1\textwidth]{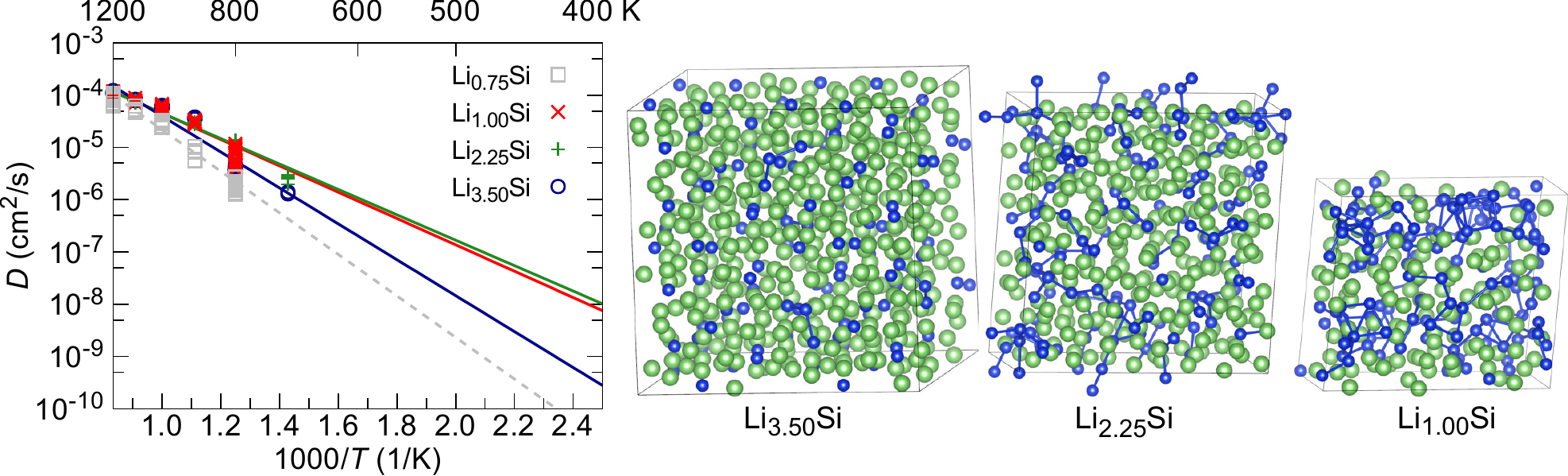}
  \caption{\label{SI-fig:arrhenius}%
    \textbf{Arrhenius extrapolation of the Li diffusivities in
      \ce{Li_{3.75-x}Si} structures to room temperature.}  The data is
    based on 5~ns long molecular dynamics simulations of structures
    containing 128~Si atoms and between 96 and 480 Li atoms
    (\ce{Li_{480-x}Si128}).  Only molecular dynamics trajectories with
    Li mean squared displacement greater than 10~\AA{}$^{2}$ were
    considered.  The corresponding room temperature diffusivities and
    activation energies are given in Table~1 in the main manuscript.
    For the sake of completeness, here the composition \ce{Li_{0.75}Si}
    is also included to show that the diffusivity decreases once the Li
    content drops below \ce{Li_{1.00}Si}.  Also shown are the structures
    of \ce{Li448Si128} (=~\ce{Li_{3.50}Si}), \ce{Li288Si128}
    (=~\ce{Li_{2.25}Si}), and \ce{Li128Si128} (=~\ce{Li_{1.00}Si}) after
    2~ns MD equilibration at $T=$~600~K.}
\end{figure}

For the diffusivity calculations, 5~ns long MD simulations of the bulk
\ce{Li_{480-x}Si128} structures (see above) at temperatures between
400~K and 1200~K were carried out.
It was found that all bulk structures generally reached thermal
equilibrium after no more than 200~ps, even compositions for which Si
segregated during the simulation.
To achieve sufficient statistical confidence, the diffusivity was
evaluated for 18~MD trajectory parts over each 450~ps after a 200~ps
equilibration phase that was discarded.
Lithium diffusivities $D$ were obtained by fitting the Einstein relation
\begin{align}
    6\,D\,t
    = \Bigl\langle
        \frac{1}{N}\Bigl|\Bigl|
           \sum_{i=1}^{N}\Delta \mathbf{r}_{i}(t)
        \Bigr|\Bigr|^{2}
      \Bigr\rangle
  \label{eq:Einstein}
\end{align}
where $t$ is the time, $N$ is the number of Li atoms in the structure,
and $\Delta\mathbf{r}_{i}$ is the displacement of atom $i$ from its
position at $t=0$.
The fits were performed using tools implemented in the pymatgen software
package~\cite{cms68-2013-314}.
Using the net displacement instead of the atomic displacement in
equation~\eqref{eq:Einstein} accounts for correlated motion of Li
ions~\cite{ncm2-2016-16002}.
Only those MD trajectories with a final net displacement larger than
10~\AA{}$^{2}$ were analyzed to guarantee sufficient Li hopping.
The activation energies for Li diffusion and the diffusivities at room
temperature were obtained from Arrhenius fits for temperatures in the
range 400--1200~K (Supplementary \textbf{Fig.~\ref{SI-fig:arrhenius}}).

\clearpage
\section{Supplementary Information}

\subsection{Validation of the Artificial Neural Network Potential}

To ensure that the artificial neural network (ANN) potential is reliable
for molecular dynamics (MD) simulations of amorphous \ce{Li_xSi}
structures with different compositions, we extensively benchmarked the
potential on relevant structures that were not used for the potential
training.

\begin{figure}[tbp]
  \centering
  \includegraphics[width=0.6\textwidth]{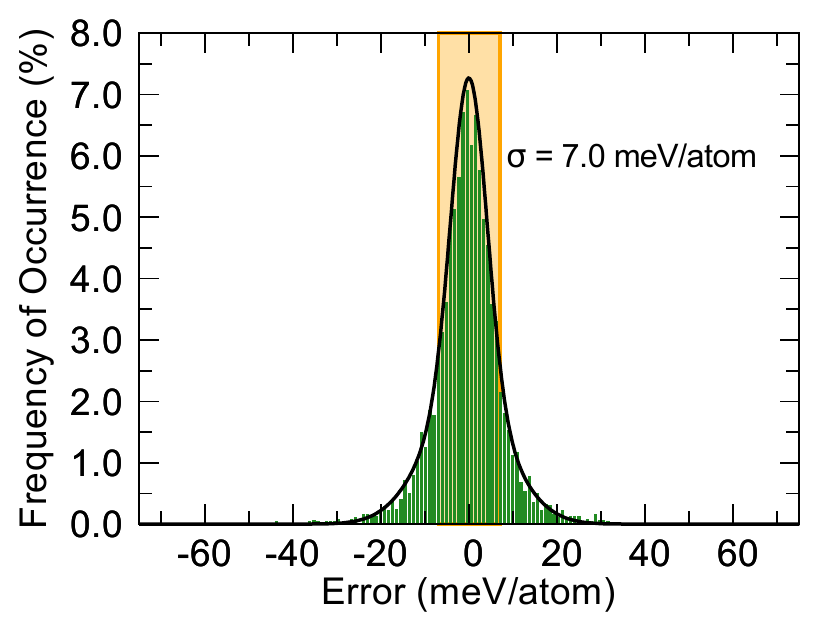}
  \caption{\label{SI-fig:ANN-error-distribution}%
    \textbf{Error distribution in the test set used for the validation
      of the ANN potential.} The test set comprises 4,516 structures
    that were randomly selected from the reference data set and were not
    used for the training of the ANN potential.  The orange region
    indicates the standard deviation of $\sim$7.0 meV/atom of the fitted
    distribution (black lines). The root mean squared error (RMSE) of
    the ANN potential energy is $\sim$7.7~meV/atom, which includes the
    tails that are not captured by the fitted distribution (the sum of
    two normal distributions). The RMSE of the training set is
    $\sim$6.3~meV/atom.}
\end{figure}

Figure~\ref{SI-fig:ANN-error-distribution} shows the distribution of errors
in energies predicted by the ANN potential relative to their DFT
references for all structures in the test set. As seen in the figure,
the error distribution is symmetric and zero-centered, indicating that
the space of the reference structures is homogeneously sampled.

Since the main objective of the present work is to understand Li
transport in amorphous LiSi, we assessed the reliability of the ANN
potential for Li diffusion.
A standard technique for the calculation of diffusion barriers in solids
is the nudged-elastic-band (NEB) method~\cite{Jonsson1998,
  jcp113-2000-9901}.

\begin{figure}[tbp]
  \centering
  \includegraphics[width=\textwidth]{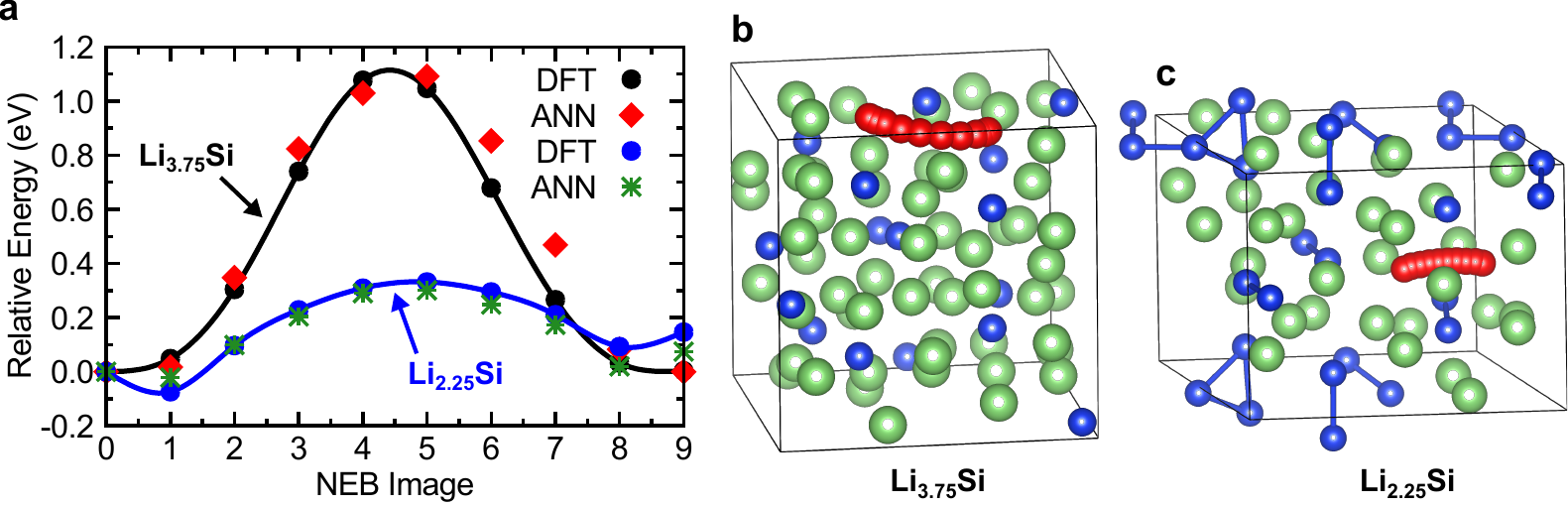}
  \caption{\label{SI-fig:NEB}%
    \textbf{Comparison of the ANN potential and DFT activation energies
      for Li diffusion in two \ce{Li_xSi} structures with different
      composition.} \textbf{a}, Energy along the minimum energy pathways
    for Li diffusion in \ce{Li_{3.75}Si} and \ce{Li_{2.25}Si} as
    calculated using the nudged elastic band method (NEB). Black and
    blue circles indicate the DFT energies for the NEB images, red
    diamonds and green stars are the corresponding ANN potential
    energies. The lines are cubic splines to guide the eye. \textbf{b}
    and \textbf{c}, Selected Li diffusion pathway in the
    \ce{Li_{3.75}Si} (= \ce{Li60Si16}) structure and in the
    \ce{Li_{2.25}Si} (\ce{Li36Si16}) structure, respectively. Si atoms
    are colored blue, Li green, and the NEB images red.  Neither the
    endpoints nor the NEB images were included in the ANN potential
    training set.}
\end{figure}

Figure~\ref{SI-fig:NEB} shows a comparison of the ANN potential and DFT
energies along the minimum energy diffusion paths for two different
\ce{Li_xSi} compositions.
The initial and final points of the diffusion path were created by
removing neighboring Li atoms from the corresponding \ce{Li_xSi}
structures.
Neither the endpoints nor the intermediate structures were included in
the ANN potential training.
As seen in the figure, the barriers predicted by the ANN potential agree
excellently with the DFT results.

NEB calculations require the optimized endpoints of the diffusion as
input.
In fully amorphous \ce{Li_xSi} structures, such as in the
\ce{Li_{1.00}Si} phase, the removal of a single Li atom results in
significant structural rearrangement which makes NEB calculations
challenging.

To obtain a benchmark for Li diffusion in such amorphous structures, we
evaluated the energy along interpolated paths in four additional
structures with different compositions.
In each case, the final structure from a 2~ns long MD trajectory at
600~K was analyzed.
The resulting diffusion paths are not minimum energy paths and are
rather high in energy, but the comparison of the ANN potential results
with the DFT energies allows assessing the reliability of the potential
for non-equilibrium structures.

\begin{figure}[tbp]
  \centering
  \includegraphics[width=0.7\textwidth]{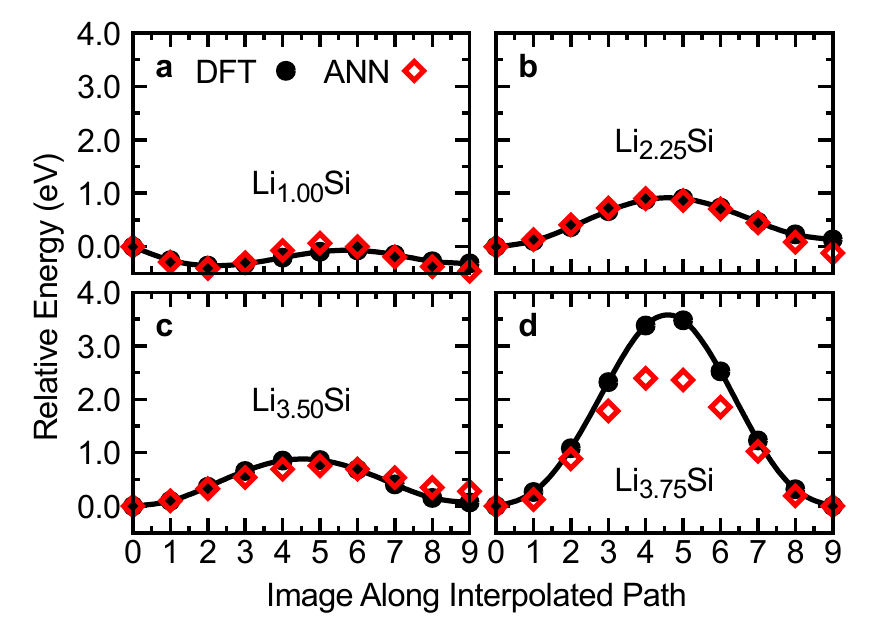}
  \caption{\label{SI-fig:static-paths}%
    \textbf{Energy along interpolated paths in \ce{Li_xSi} structures
      with four different compositions.} In contrast to the NEB paths
    shown in Figure~\ref{SI-fig:NEB}, the paths shown in this figure are
    not minimum energy diffusion paths. The DFT energies are shown as
    black circles; the corresponding ANN potential energies as red
    diamonds. }
\end{figure}

As seen in Figure~\ref{SI-fig:static-paths}, the agreement between the ANN
potential energies and DFT is very good despite the high-energy
structures along the interpolated paths.
Only for the \ce{Li_{3.75}Si} composition the ANN potential deviates
significantly from DFT for structures that are more than 2~eV above the
minimum.
However, even for the \ce{Li_{3.75}Si} composition, the ANN potential
reproduces the minimum energy path well, as shown in
Figure~\ref{SI-fig:NEB}.

\begin{figure}[tbp]
  \centering
  \includegraphics[width=\textwidth]{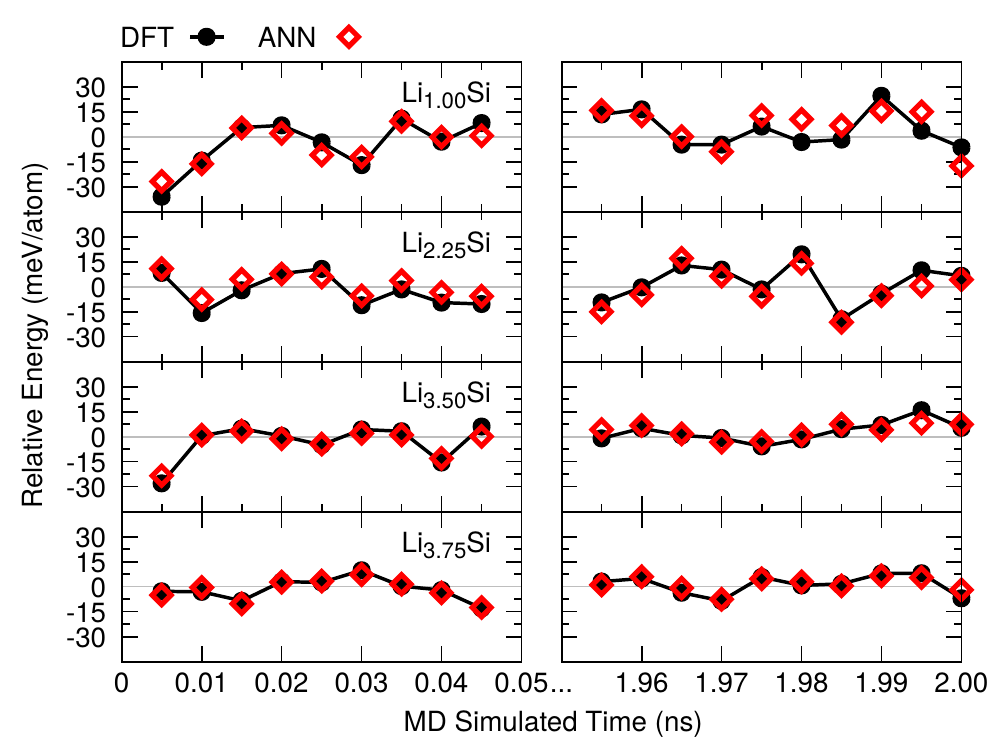}
  \caption{\label{SI-fig:MD-ANN-vs-DFT}%
    \textbf{Comparison of energies along MD trajectories: bulk
      \ce{Li15Si4} (Li\,=\, 480, 448, 288, and 128 atoms).} Energies
    during the initial (left) and final (right) 50~ps of 2~ns long MD
    trajectories are shown. The ANN potential energies are shown as red
    diamonds, and the corresponding DFT energies are black circles. }
\end{figure}

Finally, we assessed the reliability of the ANN potential for actual MD
simulations by comparing the energies of structures during the initial
and final 50 ps of MD trajectories at $T$\,=\,1000~K.
As seen in Figure~\ref{SI-fig:MD-ANN-vs-DFT}, the agreement between the ANN
potential and DFT is remarkable.
Deviations increase with decreasing Li content but never exceed
10~meV/atom.
An even closer agreement can be expected for lower temperatures that
give rise to less distorted structures during the MD simulation.
This comparison also confirms that the MD simulations with the ANN
potential do not result in trapping in artificially stabilized phases.

\begin{figure}[tbp]
  \centering
  \includegraphics[width=0.7\textwidth]{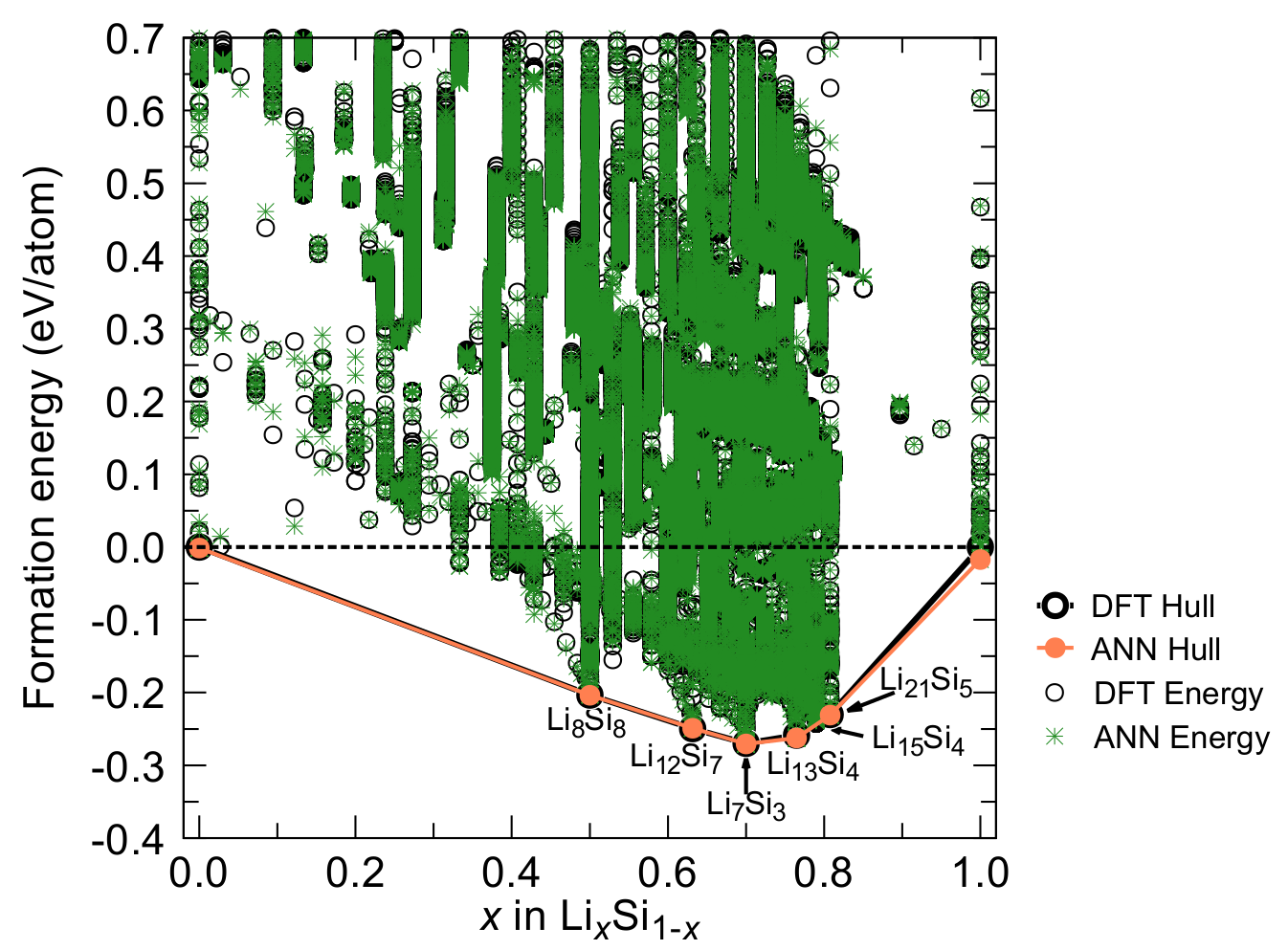}
  \caption{\label{SI-fig:hull-all-structures}%
    Comparison of the formation energies of \ce{Li_xSi_{1-x}} structures
    as predicted by the artificial neural network (ANN) potential (green
    stars) with their density-functional theory (DFT) reference energies
    (black circles).  The black and yellow lines and filled circles
    indicate the lower convex hulls of the DFT and ANN energies,
    respectively.  The ANN potential ground states are in excellent
    agreement with DFT.  Note that this plot shows the formation
    energies of structures from the entire reference data set.}
\end{figure}

\begin{figure}[tbp]
  \centering
  \includegraphics[width=\singlecol]{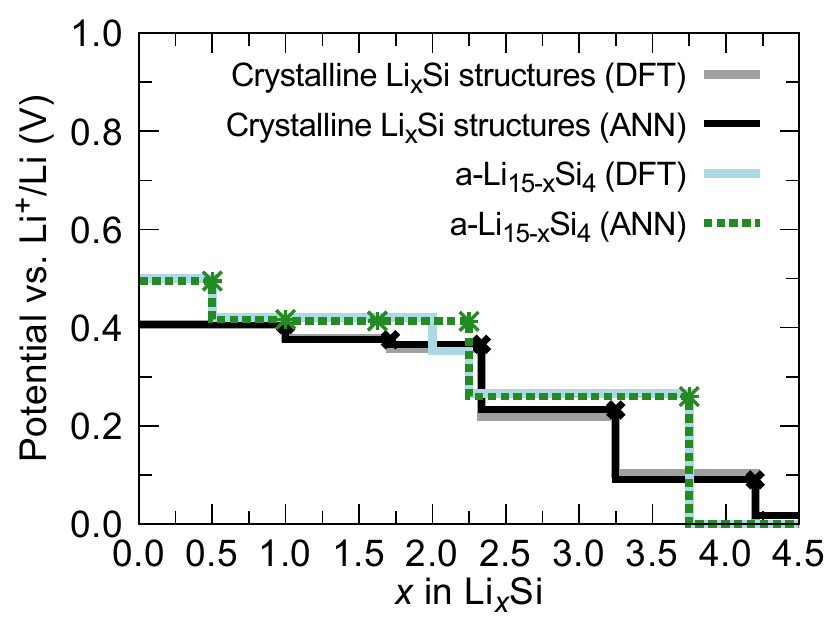}
  \caption{\label{SI-fig:voltage-profile}%
    Comparison of computational 0~K voltage profiles of the
    thermodynamic ground state phases of the \ce{LiSi} alloy and of
    amorphous a-\ce{Li_{15-x}Si4} phases predicted by DFT and the ANN
    potential.  As seen in the figure, the voltage profile predicted by
    the ANN potential is nearly identical to the DFT voltage profile,
    further confirming the accuracy of the potential.  Note that the
    crystalline \ce{Li_xSi} structures were included in the ANN
    potential training set.}
\end{figure}

{\linespread{1.4}
\begin{figure}[tbp]
  \centering
  \includegraphics[width=\linewidth]{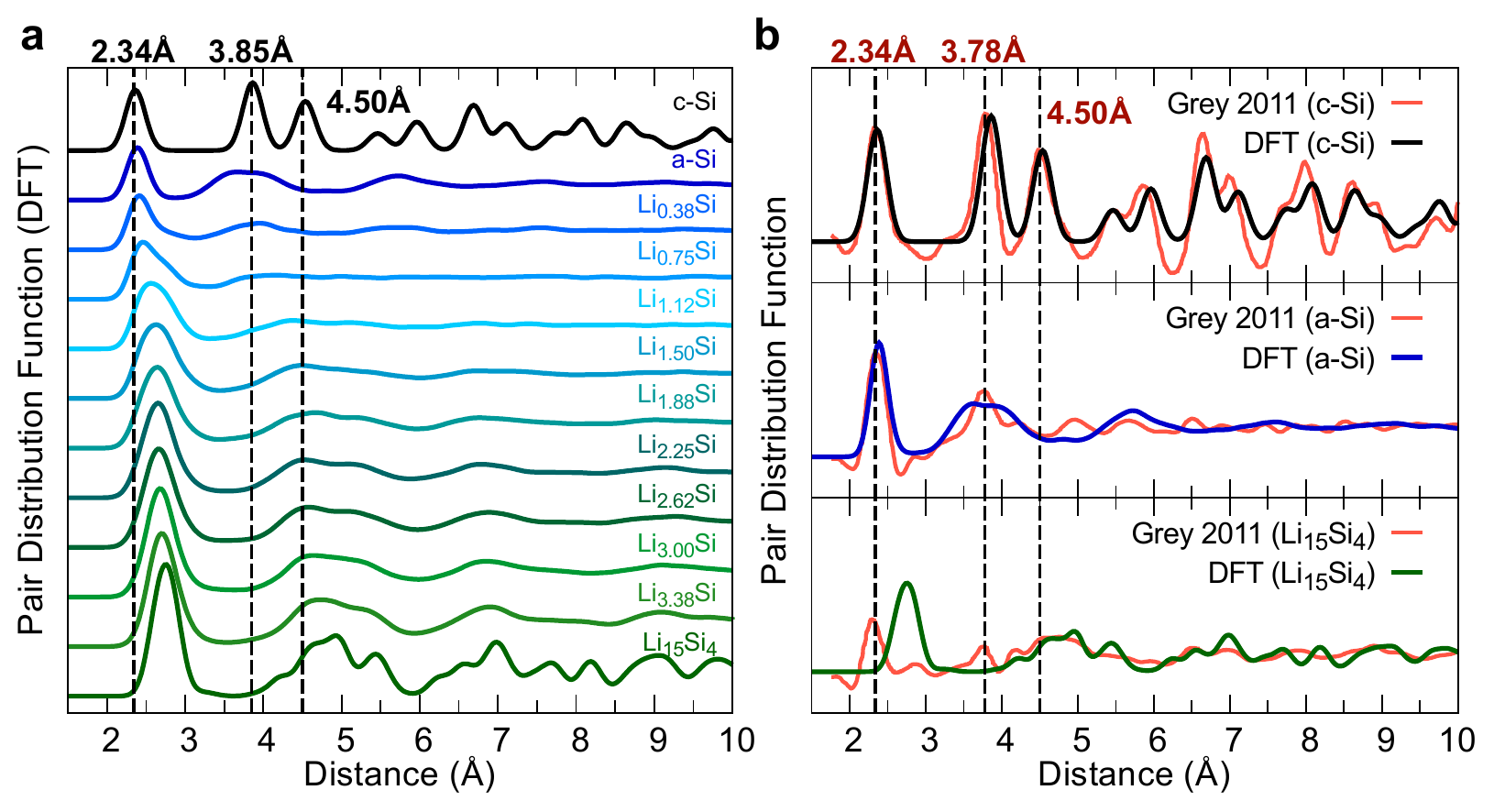}
  \caption{\label{SI-fig:pdf-Li480Si128}%
    \textbf{Pair distribution functions for crystalline and amorphous
      \ce{LiSi}.} \textbf{a}, Calculated pair distribution functions
    (PDFs) of crystalline silicon (black) and c-\ce{Li15Si4} (green) and
    amorphous a-\ce{Li_{15-x}Si4} structures with intermediate
    compositions (shades of blue).
    The PDFs of the a-\ce{Li_{15-x}Si4} compositions were obtained from
    DFT optimized structures based on the \ce{Li480Si128} composition.
    At each composition, the lowest energy atomic configuration among the
    corresponding \ce{Li_{480-x}Si128} structures with DFT-optimized
    geometry and cell parameters was considered.
    The \ce{Li15Si4} crystal structure was taken from the international
    crystal structure database (ICSD ID:~159397) and optimized with DFT for
    consistency~\cite{jap102-2007-53704, jcics23-1983-66}.
    The atomic scattering factors were approximated by the atomic mass.
    All PDFs were convoluted with Gaussian functions with a full width
    at half maximum (FWHM) of 0.3~\AA{} to simulate experimental
    resolution.
    \textbf{b}, Comparison of the computed PDFs for c-\ce{Si},
    a-\ce{Si}, and c-\ce{Li15Si4} with measurements from
    reference~\citenum{jacs133-2011-503}.
    Both, the predicted intensities and peak positions of the computed PDF
    for crystalline silicon are in excellent agreement with the measured
    reference (top panel), indicating that the DFT lattice parameters
    are close to the real values.
    The measured PDFs for the \ce{Li15Si4} structure (bottom panel)
    contains a number of intensities that are not present in the
    computed PDF.
    Most notably, the first and second neighbor peaks of c-\ce{Si} are
    present in the X-ray \ce{Li15Si4} PDF showing that unlithiated
    silicon was still present in the measured sample.
    Similarly, all computed intensities of the amorphous silicon
    structure obtained from complete delithiation of the \ce{Li480Si128}
    structure are present in the measured PDF, but the measured PDF
    contains additional correlations that can be attributed either to
    remaining c-\ce{Si} or to residual lithium in the material (peaks at
    5.0~\AA{} and 6.5~\AA{}).}
\end{figure}
}

\begin{figure}[tbp]
  \centering
  \includegraphics[width=1.2\singlecol]{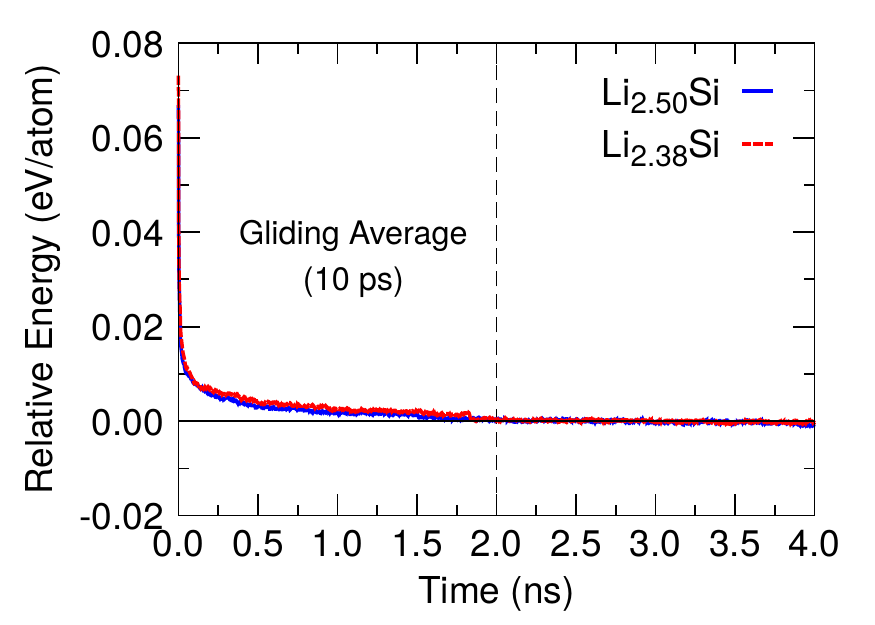}
  \caption{\label{SI-fig:MD-equi}%
    \textbf{The nanoparticle structure have reached thermal equilibrium
      after 2~ns long molecular dynamics simulations at 500~K.}  Gliding
    average (10~ps) of the potential energy during molecular dynamics
    simulation of two nanoparticle models with compositions
    \ce{Li6429Si2571} (=~\ce{Li_{2.50}Si}, blue line) and
    \ce{Li6336Si2664} (=~\ce{Li_{2.38}Si}, red dashed line) at 500~K.
    After around 2~ns the potential energy is within 2~meV/atom of the
    final structure, indicating that the system has reached thermal
    equilibrium.}
\end{figure}

\begin{figure}[tbp]
  \centering
  \includegraphics[width=1.1\singlecol]{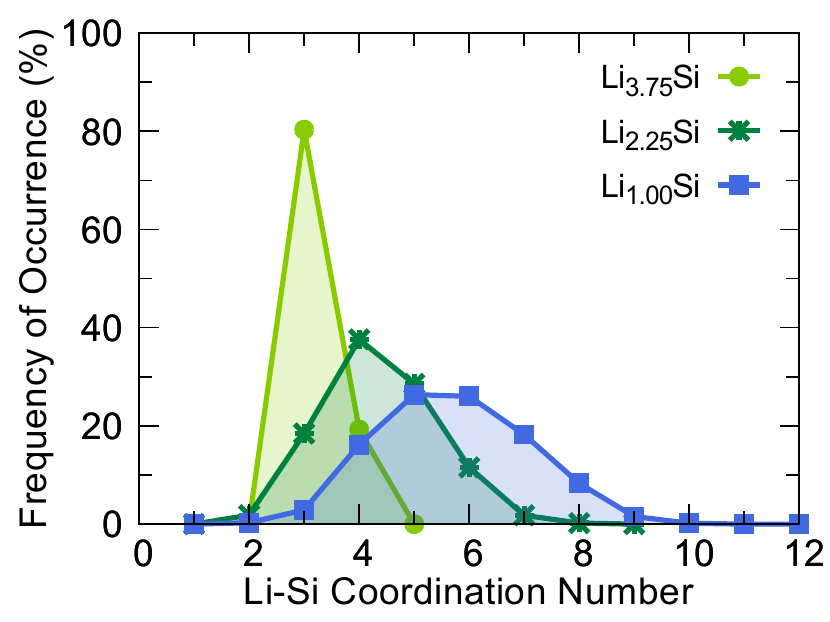}
  \caption{\label{SI-fig:cn-vs-composition}%
    \textbf{On average Li is coordinated by more Si atoms at more
      delithiated compositions but the probability of low coordinated Li
      atoms also increases.}  Number of Si atoms within a range of
    3.5~\AA{} from each Li atom during molecular dynamics simulations at
    600~K over 5~ns.  In the crystalline \ce{Li_{3.75}Si}
    (=~\ce{Li15Si4}) structure, a Li atom is coordinated either by
    3~or~4 Si atoms with around 99\% probability.}
\end{figure}

\begin{figure}[tbp]
  \centering
  \includegraphics[width=1.1\singlecol]{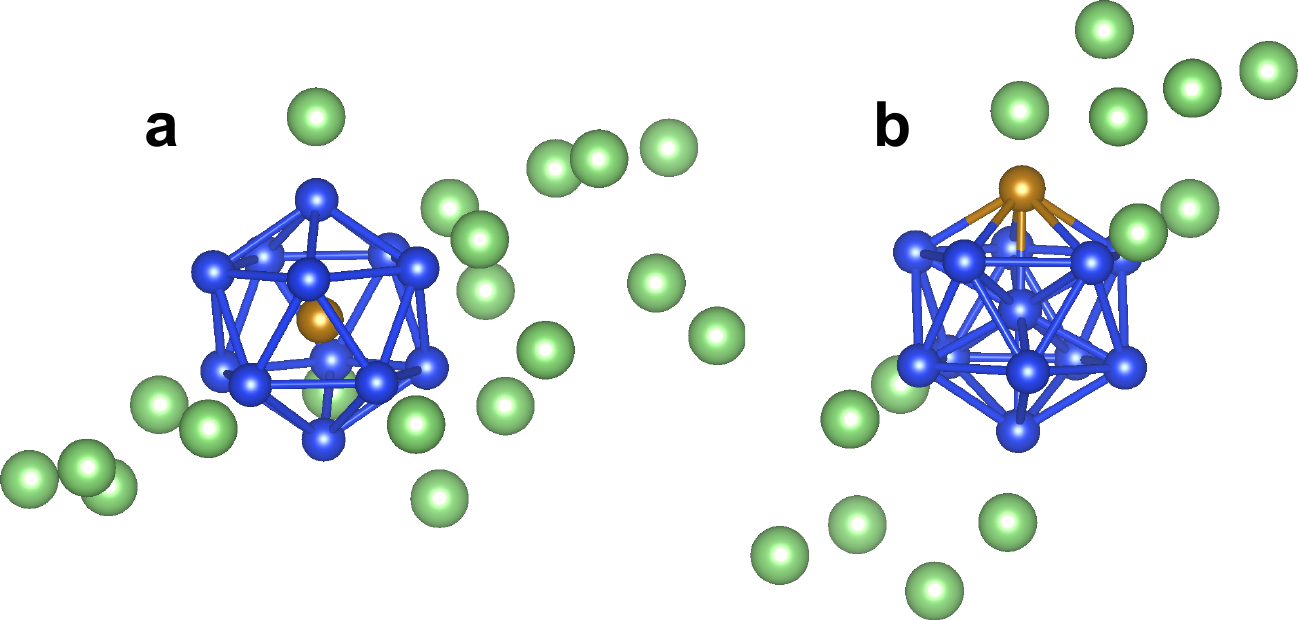}
  \caption{\label{SI-fig:ideal-doped-clusters}%
    \textbf{Initial ideal icosahedral structures of the doped cluster
      models.}  Initial structure of the \textbf{a}, \emph{core}-doped
    and \textbf{b}, \emph{shell}-doped cluster models used to estimate
    the potential for Si clustering.  Dopants that favor core-doping
    over shell-doping and bind strongly with Si are expected to promote
    Si clustering.  Li atoms are colored green, Si is blue, and the
    dopant is brown.}
\end{figure}

\begin{figure}[tbp]
  \centering
  \includegraphics[width=0.8\textwidth]{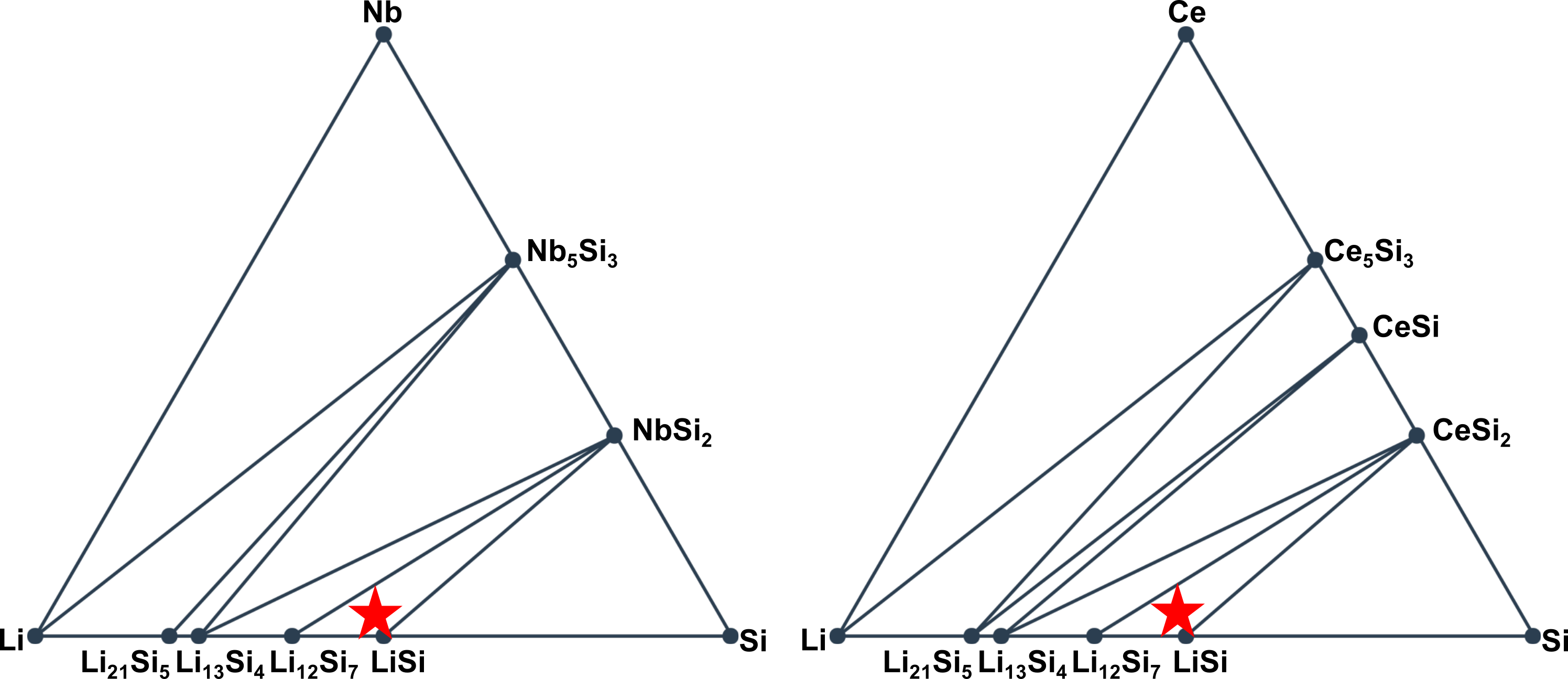}
  \caption{\label{SI-fig:PD-Nb-Si-Ce-Si}%
    \textbf{Ab initio phase diagrams of Nb and Ce compounds with Li and
      Si.}  The phase diagrams were generated based on the ab initio
    calculations of inorganic materials provided by the \emph{Material
      Project} database~\cite{aplm1-2013-011002}.  For both Nb and Ce
    compounds with Si are database but no compounds with Li.  The red
    star indicates the \ce{Li13Si12X} compositions of the investigated
    doped cluster models (Figure~\ref{SI-fig:ideal-doped-clusters}). }
\end{figure}

\begin{figure}[tbp]
  \centering
  \includegraphics[width=\textwidth]{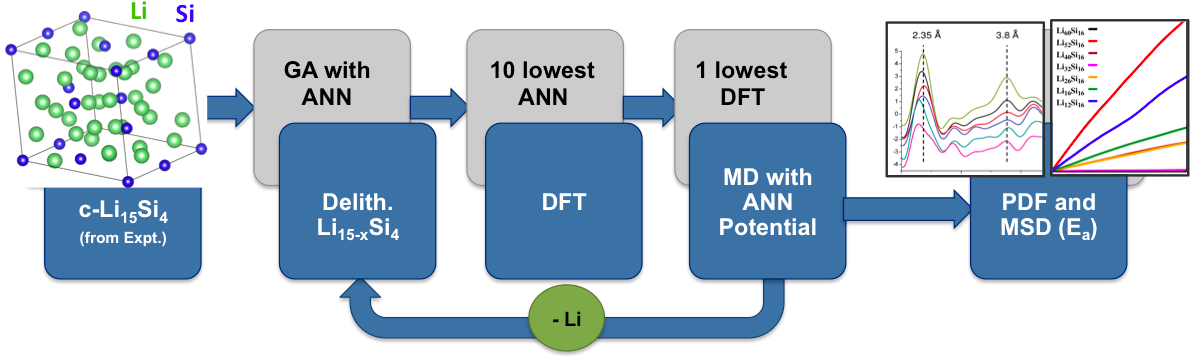}
  \caption{\label{SI-fig:GA-schematic}%
    Schematic of the structure-determination protocol used for the
    sampling of bulk \ce{Li_{15-x}Si_{4}} structures.  Different
    supercells of the experimental c-\ce{Li15Si4} structure with
    compositions up to \ce{Li480Si128} were used as starting points.
    From these structures, Li atoms were sequentially removed, and at
    each delithiated composition the lithium-vacancy ordering was
    determined with a genetic algorithm (GA).  The geometry and cell
    parameters of the 10 lowest-energy configurations were subsequently
    optimized with DFT.  The most stable structures at each composition
    were used as input for MD simulations.}
\end{figure}

\begin{figure}[tbp]
  \centering
  \includegraphics[width=\textwidth]{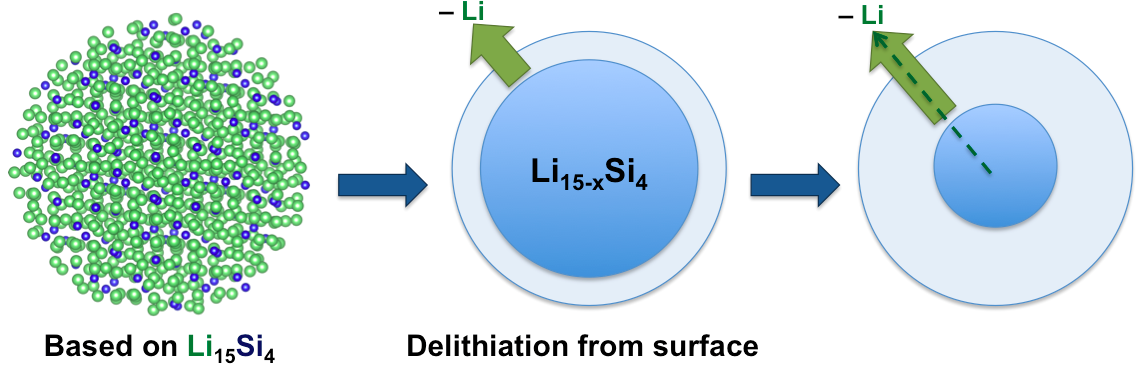}
  \caption{\label{SI-fig:NP-Delith}%
    Schematic of the protocol used for the delithiation of \ce{LiSi}
    nanoparticles.  At each step, Li atoms were removed from the surface
    of the particle, followed by a molecular dynamics simulation at
    500~K over 4~ns.}
\end{figure}

\clearpage
\subsection{Supplementary Tables}

\begin{table}[H]
  \centering
  \caption{\label{SI-tab:diffusivity}%
    Measured room-temperature Li diffusivity as reported in the
    literature.  Experimentally, the Li diffusivity has been obtained
    from electrochemical impedance spectroscopy (EIS), cyclic
    voltammetry (CV), galvanostatic intermittent titration (GITT), and
    by potentiostatic intermittent titration (PITT).}
  \vspace*{5mm}
  \renewcommand{\arraystretch}{1.0}
  \begin{tabular}{ccc}
    \hline\hline
    \multicolumn{1}{c}{\bfseries $D$ (cm$^{2}$/s)}
    & \multicolumn{1}{c}{\bfseries Method}
    & \multicolumn{1}{c}{\bfseries Reference} \\
    \hline
    10$^{-10}$ & EIS & \cite{jpcc113-2009-11390} \\
    10$^{-12}$ & CV, EIS, GITT & \cite{ssi180-2009-222} \\
    10$^{-14}$ & EIS, PITT & \cite{mcp120-2010-421} \\
    10$^{-14}$--10$^{-13}$ & PITT & \cite{jpcc116-2012-1472} \\
    \hline\hline
  \end{tabular}
\end{table}

\end{document}